\documentstyle[aps,prl,twocolumn]{revtex}
\input epsf
\begin{document}

\title{Driving-voltage-induced mechanical force oscillations in metal quantum
point contacts} 
\author{Alexandre M. Zagoskin}
\address{Physics and Astronomy Department, The University
of British Columbia, 6224 Agricultural Rd.,
Vancouver, B.C., V6T 1Z1, Canada}
\maketitle
\thanks{Email zagoskin@physics.ubc.ca}

\begin{abstract}
We predict that the mesoscopic tensile force 
fluctuations in metal  quantum point contacts  (nanowires) 
 arise as a result of finite electric voltage on the  
contact. 
They are due to reconfiguration of the  electronic subsystem
and are correlated with the nonlinearities of the current-voltage
characteristics of the contact. The observation of the effect would  
directly confirm
 the recently suggested "free-electron" mechanism of mesoscopic force fluctuations observed
in  nanowires under deformation.

The related magnetic susceptibility fluctuations 
and role of  topology of the wire cross section
are discussed as well.

   \end{abstract}

\begin{figure}
\epsfxsize=8 cm
\epsfbox{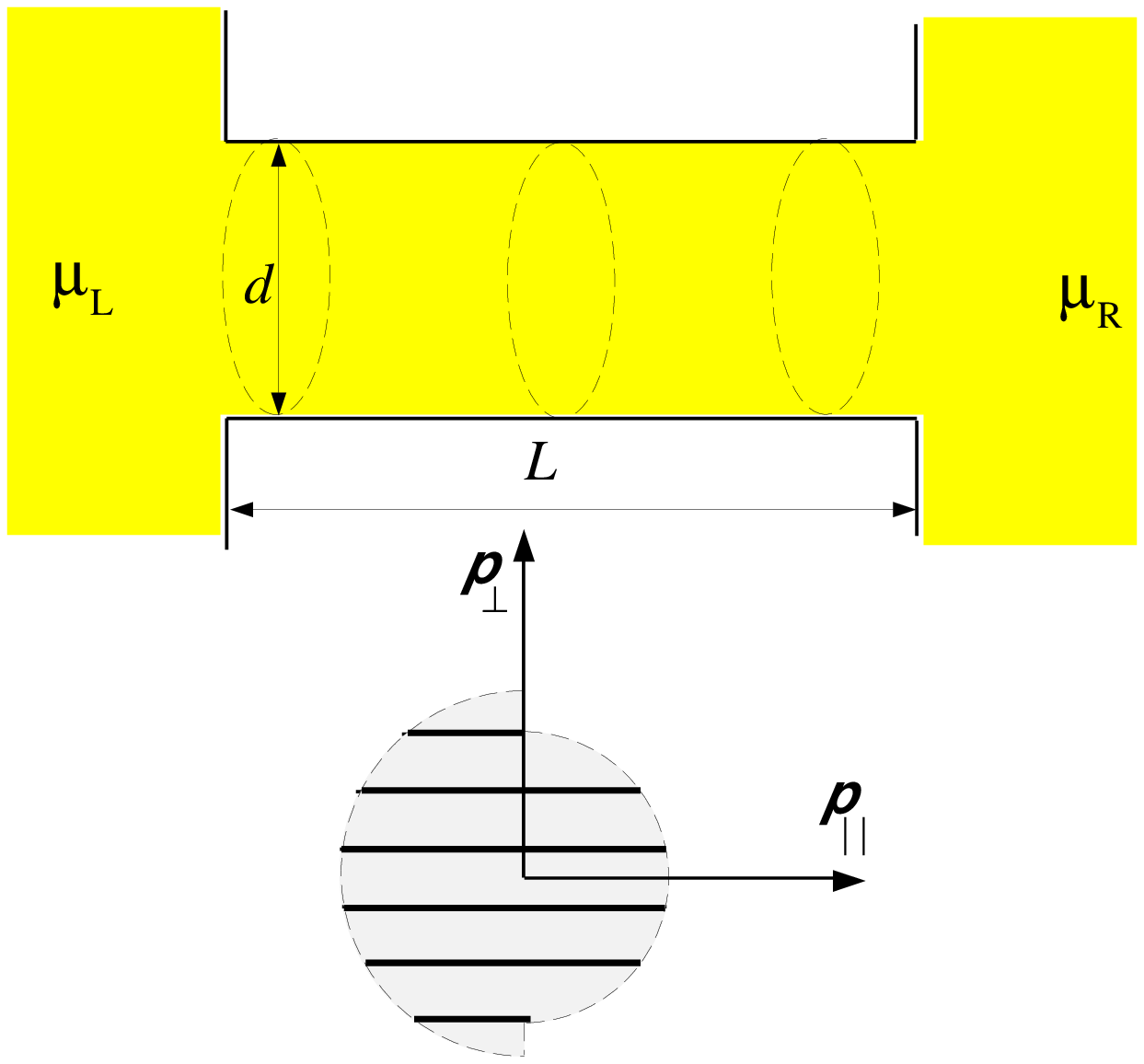}
\caption{Schematic view of a metal quantum point contact (nanowire)
and distribution function of right- and left-moving electrons
in the wire
at finite driving voltage. Solid lines
correspond to quantized values of transverse momentum.}\label{fig0}
\end{figure}

 The quantization of electrical conductance of 3D quantum point contacts (QPC) 
 \cite{777}  has been
observed in  a variety of metal  mesoscopic contacts\cite{1}.
The simplest model of 
the 3D QPC is a "nanowire'' of length $L$ and dia\-me\-ter $d 
\sim  \lambda_F$,
connecting two bulk conducting reservoirs, where $\lambda_F$ is the Fermi wavelength in the system (see Fig.~\ref{fig0}),
and the underlying atomic structure is the positively
charge "jellium".  It is quantized character of transverse motion of electrons   that is revealed in conductance quantization 
as a function of $d$. Since the applied driving voltage changes the population 
of occupied subbands,  nonlinear current - 
voltage dependence 
quantum point contacts, 
was predicted\cite{GK,Zagoskin90,IDP,BSL}. The theory is in a  
good agreement with experiment on 2D and 3D systems\cite{Pepper}. 
   On the other hand,  the   nonlinear 
conductivity  of   3D bismuth QPC\cite{dremov} shows only
qualitative agreement with the theory, while measurements on
3D gold QPCs\cite{yasuda} did not show the predicted
type of nonlinearities altogether\cite{buttiker}.

The experiments on metal QPCs under deformation showed that 
 the mechanical stress in the wire fluctuates as a function of
its elongation, and the fluctuations are  correlated with 
the conductance jumps\cite{8}.  This seemed to require a more 
complex mechanism, and the often-used explanation of these phenomena
 invokes the atomic rearrangement processes, and is supported
by molecular dynamics calculations (e.g., \cite{9}). 

 Recently, an elegant alternative   explanation of the force fluctuations 
was suggested,
 based on the reaction of
the free electrons  to the mechanical deformation of the contact 
region\cite{12}. The role
of the atomic structure of the wire was again reduced to providing 
a ``jellium" background. 
In this "free-electron" model, the longitudinal
force, being the coordinate 
derivative of the thermodynamic potential, is sensitive
to the positions and occupancy of electronic subbands in the wire.  The latter depends on the shape of the cross section and 
on the elongation of the wire (assumed to
take place at a constant volume).     The positions of cusps in $F$ as a function of the elongation would naturally 
coincide with those of the conductance steps.

In this paper we show that  
the "free-electron" mechanism of tensile force fluctuations will lead to 
related
  effects:  mechanical
 force  and magnetic susceptibility  fluctuations in a nanowire  as a function
of applied driving voltage at the same values of $eU$,
as the features  of  the differential conductance, $G_{\rm diff}(eU)$. 
  Investigation of these effects can be done using existing experimental techniques and would
  provide an independent test of the mechanism suggested in \cite{12},
and  confirm  the decisive role
of electronic subsystem in determining both transport and mechanic  
properties of metal quantum contacts.

We   start  from the expression for the grand potential of
the electronic subsystem at zero temperature  and voltage
for a wire of uniform cross section (Blom et al.\cite{12}):
\begin{eqnarray}
\Omega(E_F) =  -\frac{4}{3}L\sqrt{\frac{2m^*}{\pi^2\hbar^2}} \sum_n\left(E_F-E_n(L,V)\right)^{3/2}\nonumber\\
\times\theta(E_F-E_n(L,V)). \label{O}
\end{eqnarray}
 Here $m^*$ is the
effective mass of an electron, $\theta(x)$ is the Heaviside step function,
and
$E_n$ is the energy of $n$th electronic transverse mode  
in the wire, which depends on
  the wire length, $L$, and volume, $V$. 
We   assume that the length of the wire is  much larger than 
its diameter $d$, which allows us to set the electrical potential to zero
in the wire. (Due to screening in metal 
the effects of finite bias will be felt only at  distances 
$\approx d$ from the ends, see \cite{Zagoskin90} and references therein.)

For the sake of simplicity, we will neglect both elastic and inelastic 
scattering in the nanowire. Their contributions are of order $(L,d)/l_s$, where
$l_s$ is the corresdponding scattering length \cite{KOS}. The condition
 $(L,d)/l_s \ll 1$ is satisfied both for the elastic  and  electron-phonon scattering. 
In the latter case we use the estimate 
$l_{e-ph} \sim \hbar v_F/\lambda\omega_D$, where $\omega_D$ is Debye frequency,
and $\lambda<1$ is the electron-phonon coupling constant.
The electron-electron  scattering length can be estimated as 
  $l_{e-e} \sim \hbar v_F/\epsilon_F (\epsilon_F/eV)^2$,
the bias $eV$ playing the role of effective temperature. For the effects of 
electron-electron scattering to be small we need $d/l_{e-e}\ll 1$
(because the longitudinal momentum conservation and transverse quantization
 in the electronic subsystem
suppress the electron-electron scattering inside the nanowire). 
 The corresponding restriction on the applied voltage is
$
eV < \epsilon_F/\sqrt{N_{\perp}},
$
where $N_{\perp}\sim d/\lambda_F$ is the number of transverse channels in
the quantum contact. This condition is compatible with bias being
of the same order as the interlevel spacing in the contact, 
$
\Delta\epsilon \sim \epsilon_F/N_{\perp},
$
 which is necessary for the observation of
nonlinear effects discussed in this paper.

Under the above assumptions, we can consider
right-moving and left-moving electrons as two inde\-pen\-dent subsystems, with
chemical potentials $\mu_L$ and $\mu_R$ respectively\cite{KOS}
(Fig.~\ref{fig0}):
\begin{equation}
\mu_L = E_F - (1-\beta)eU; \:\:\:\: \mu_R = E_F + \beta eU.
\end{equation}
The grand potential thus becomes 
\begin{eqnarray}
 \Omega_{eU}  = \frac{1}{2}\left(\Omega(E_F + \beta eU)
 + \Omega(E_F - (1-\beta)eU)\right).
 \end{eqnarray}
Parameter $\beta$
determines the asymmetry of the voltage drop on the contact. (Usually  
symmetric voltage drop is assumed ($\beta = 1/2$), in which case
the differential conductance is always a multiple of $\frac{1}{2}G_Q$.
\cite{GK} 
Generally 
$\beta$ can deviate from $1/2$ (\cite{Pepper}) and be voltage-de\-pen\-dent.)
It should be in principle determined self-consistently by solving
the electrostatical problem for the wire and its surroundings
at   given $eU$\cite{buttiker}. We consider here two limiting cases:
(a)  symmetric voltage drop, $\beta = 1/2$, and
(b) perfect screening. In the latter case $\beta(eU)$
is determined from the condition of no charge accumulation in
the wire,
\begin{equation}
N(E_F+\beta eU)+N(E_F-(1-\beta)eU) = 2 N(0), \label{N const}
\end{equation}
where
\begin{eqnarray}
N(E) = 2 L \sqrt{\frac{2 m^*}{\pi^2\hbar^2}} \sum_n \theta(E - E_n)\sqrt{E - E_n}.  
\end{eqnarray}
  
It is easy to see, that   the differential conductance of the system  is given by
\begin{eqnarray}
G_{\rm diff}(eU) = \frac{dI}{dU} = \frac{d}{dU} \frac{2e}{h} 
\int_{E_F-(1-\beta)eU}^{E_F+\beta eU} dE \sum_n\theta(E-E_n) \nonumber\\
=   G_Q \sum_n \{ \beta\theta(E_F+\beta eU-E_n) \nonumber\\
+ (1-\beta)\theta(E_F-(1-\beta)eU-E_n)   \label{GGG}\\
  + U \frac{d\beta}{dU} 
 (\theta(E_F+\beta eU-E_n) -\theta(E_F-(1-\beta)eU-E_n)) \},
\nonumber
\end{eqnarray}
where  $G_Q =  2e^2/h$
is the unit quantum conductance.  
Therefore $G_{\rm diff}(eU)$
shows a structure   at the voltages when consequent
  transverse energy levels enter the current-carrying interval  
  $[E_F-(1-\beta)eU, E_F+\beta eU].$ In the limit of perfect screening
$\beta(eU)$ was found numerically for both models we considered: the wire
of square cross-section $d\times d$, and the roun wire of diameter $d$. 
 
\begin{figure} 
\epsfxsize=8 cm
\epsfbox{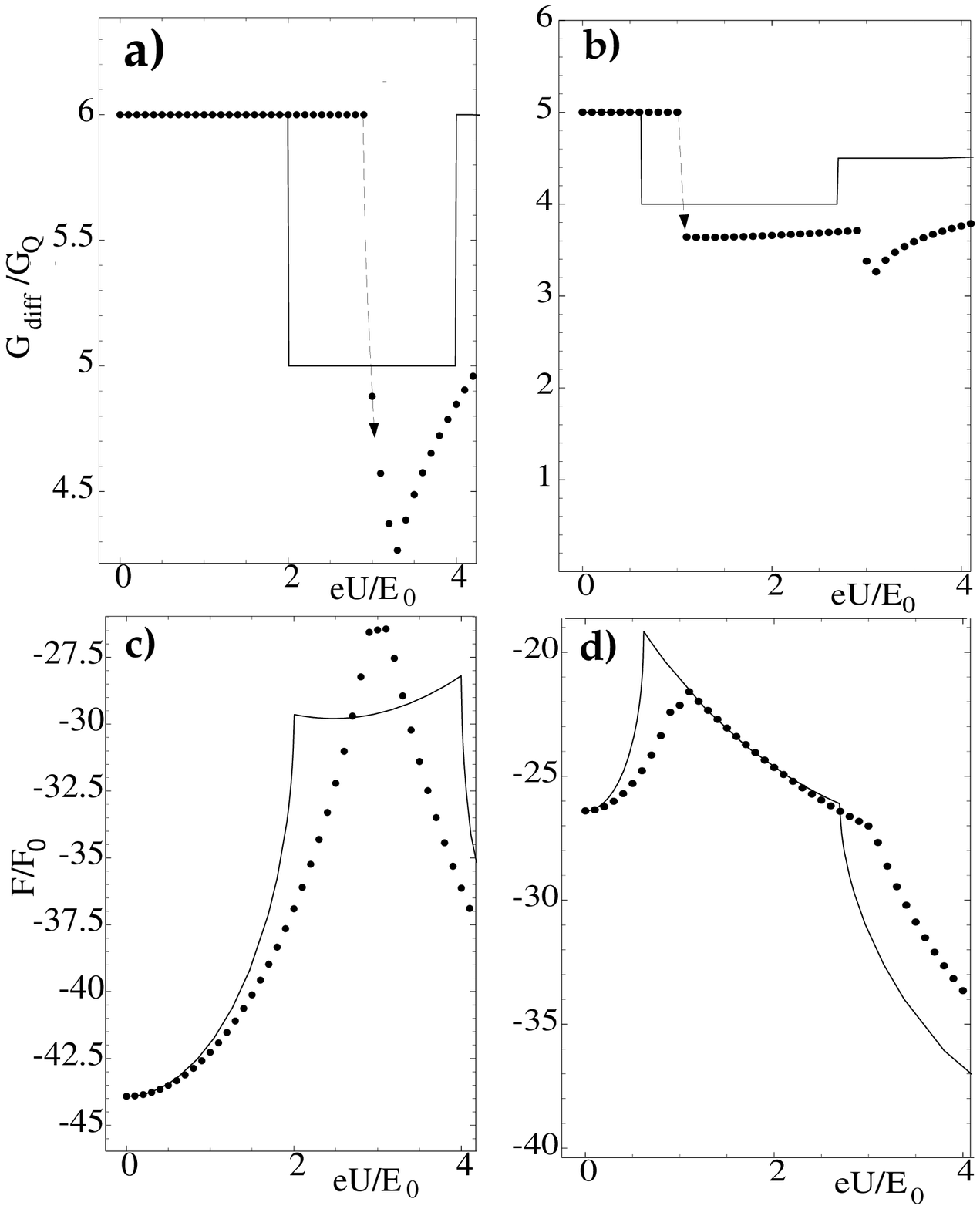}
\caption{Differential conductance and force  fluctuations 
vs. applied voltage in a nanowire of square cross section $d\times d$  
(a,c) or circular cross section of diameter $d$ (b,d).
 The force and bias are measured in units of
$F_0 = \pi^2\hbar^2/(2m^*d^3)$ and
$E_0 =  \pi^2\hbar^2/(2m^*d^2)$
respectively. The Fermi energy is
 $E_F = 11 E_0$. 
Solid line:  symmetric voltage drop
($\beta = 0.5$). Dots: perfect screening.
 }\label{fig2}
 \end{figure}

The quantized levels in the wire are given by  
\begin{eqnarray}
 E_{mn}  = \left\{ \begin{array}{ll} E_0(m^2+n^2) & {\rm (square)};\\
\frac{4 E_0}{\pi^2}
 \gamma_{mn}^2 & {\rm (round)}.  
\end{array}\right.
\label{tildalevels0}
\end{eqnarray}
 Here $E_0 =  \pi^2\hbar^2/(2m^*d^2)$, and   $\gamma_{mn}$  is denotes the 
$n$th positive
zero of the Bessel function $J_m(z)$; 
  $m$ is the magnetic quantum number.  
The mechanical force is given by
\begin{eqnarray}
F(u) = -\left(\frac{\partial \Omega}{\partial L}\right)_V  
= F_0
\sum_{mn} \left[ f\left( \epsilon_F + 
\beta(u) u ; \epsilon_{mn} \right)  \right.\nonumber\\
\left.+ f\left( \epsilon_F 
- (1-\beta(u))u; \epsilon_{mn}\right)\right],
\end{eqnarray}
where 
$\epsilon_F =  E_F/E_0; \:\: u = \ eU/E_0; \:\: 
\epsilon_{mn} =  E_{mn}/E_0; \:\:
F_0 = \pi^2\hbar^2/(2m^*d^3),
$
and
$
f\left(x;y\right) =     \left((2/3)(x - y)^{3/2} -   (x - y)^{1/2} y
\right) 
\theta\left(x - y\right).$
The nonlinear dependence $F(eU)$ and $G_{\rm diff}(eU)$ 
in both limiting cases is shown in
Fig.~\ref{fig2}. The   nonlinear conductance is strongly de\-pen\-dent on the 
character of screening in the wire.   On the other hand,  the qualitative
character of the force fluctuations is the same,
and mechanical force and differential conductance 
still show singularities at the same applied voltages in both
limits\cite{NOTE}. 
 The absolute magnitude of force fluctuations for $d\sim 1$nm is of order
$1$ nN, in agreement with earlier results\cite{8,12}.

Another way of changing positions of quantized levels, and thus
the properties of the contact, is by applying longitudinal magnetic 
field. The characteristic field sweep scale,
corresponding to interlevel spacing, is though
$\sim \Phi_0/d^2$, where $\Phi_0 = hc/e$ is magnetic flux quantum.\cite{777,BSL,ZagoskinKulikJPCM} For a metal QPC
  with $d \sim 1$~nm this yields unrealistic
fields of order $10^3$ T. This means that in metal contacts,
appreciable dependence of contact's properties on the magnetic field
can take place only when the Fermi level is already very close to one of
the quantized  energy levels. This can be achieved, e.g., by mechanical
deformation of the contact, or by applying finite driving voltage. We will
concentrate on the latter possibility, which is reversible and promises better
opportunities for the necessary fine tuning.  

The magnetic field can be taken 
 into account in the perturbation theory\cite{Bogadze}, valid in 
   the limit 
of weak field (large cyclotron radius, $r_c \gg d$)\cite{FN}one finds for the quantized transverse levels 
\begin{eqnarray}
 \tilde{E}_{mn}^s(\eta ) = \frac{4 E_0}{\pi^2}
 \left(\gamma_{mn}^2 + 2m\eta  +  4\frac{m^*}{m_0}g\eta s \right)
+ O(\eta^2).
  \label{tildalevels}
\end{eqnarray}
Here $\eta$ is the    magnetic field measured in units of
$H_0 = (hc/e)/(\pi d^2/4).$
Since $\eta\ll 1$, we keep in (\ref{tildalevels}) only linear in $\eta$
terms, including the spin splitting (the last term in the parentheses).
Here  $s = \pm 1$ 
is the projection of spin, $g$ is the $g$-factor, and $m_0$ is the
free electron mass.

The differential conductance   and force fluctuations
are thus given by
\begin{eqnarray}
G_{\rm diff}(\eta,u) = \frac{1}{2}
G_Q\! \sum_{s=\pm 1}\!\sum_{n=1}^{\infty}\!\sum_{m=-\infty}^{\infty}\! 
\left\{\theta\left(\!\epsilon_F \!+ \beta u \!- \tilde{\epsilon}_{mn}^{s}
(\eta )\right)  \right.\nonumber\\ 
\left.\!\!\!\!\!\!\!+ \theta\left(\epsilon_F \!- (1\!-\!\beta)u \!- \tilde{\epsilon}_{mn}^{s}
(\eta )\right) \right. 
  + u\frac{d\beta}{du}   \\
 \times \left[\theta\left(\epsilon_F \!+ \!\beta u - \!\tilde{\epsilon}_{mn}^{s}
(\eta)\right) \left. \!-  \theta\left(\epsilon_F \!-\! (1-\beta)u - 
\!\tilde{\epsilon}_{mn}^{s}
(\eta)\right)
\right]\right\},\nonumber
\end{eqnarray}
\begin{eqnarray}
F(\eta,u) = \frac{1}{2} F_0 
\sum_{s=\pm 1}\sum_{n=1}^{\infty}\sum_{m=-\infty}^{\infty}  \left\{ f\left(\epsilon_F+\beta u; \tilde{\epsilon}_{mn}^s(\eta)\right) \right.\label{RForce}
\\
\left.
+ 
f\left(\epsilon_F-(1-\beta)u; \tilde{\epsilon}_{mn}^s(\eta)\right)\right\},\nonumber
\end{eqnarray}
where 
$
\tilde{\epsilon}_{mn} \equiv  \tilde{E}_{mn}/E_0$.
The factors of one half before $G_Q, F_0$
reflect the spin splitting in the magnetic
field of previously degenerate energy levels. 
 
The magnetization of the wire is
\begin{eqnarray}
{\cal M}(\eta ,u)
 = -\frac{1}{V}\left(\frac{\partial\Omega}{\partial H}\right)_{V,T} 
= \frac{m_0}{m^*} \mu_B
\sum_{s,m,n}\left(-\frac{\partial \tilde{\epsilon}_{mn}^s(\eta)}{\partial\eta}\right) \nonumber\\
 \!\!\!\!\!\!\!\times\left\{\left(\epsilon_F+u/2-\tilde{\epsilon}
_{mn}^s(\eta)\right)^{1/2}
\theta\left(\epsilon_F+u/2-\tilde{\epsilon}_{mn}^s(\eta )\right)
\right.  \\
\left. +  \left(\epsilon_F-u/2-\tilde{\epsilon}_{mn}^s(\eta)\right)^{1/2}
\theta\left(\epsilon_F-u/2-\tilde{\epsilon}_{mn}^s(\eta)\right)
\right\},
\nonumber
\end{eqnarray}
where $\mu_B = e\hbar/(2m_0c)$ is the Bohr magneton.

The effects of the applied weak
magnetic field are described by   magnetoconductance coefficient,
$\sigma(u) \equiv  \left( \partial G_{\rm diff}/\partial H\right)_{V,T; H=0}$, 
 magnetotension
  coefficient, $\Upsilon(u) \equiv \left(\partial F/\partial H
\right)_{V,T; H=0},$  and
the magnetic susceptibility,
$\chi(u) = \left(\partial {\cal M}/\partial H\right)_{V,T; H=0}.$ 

Keeping only the singular terms, we find the following expressions:
\begin{eqnarray}
\!\!\!\!\!\!\!\sigma(eU) = -\frac{1}{2}
G_Q \sum_{s,n,m}\left(\frac{\partial\tilde{\epsilon}_{mn}^{s}(\eta)}
{\partial\eta}\right)_{\eta=0}\\
\!\!\!\!\left\{\delta\left(\epsilon_F\!+\beta\!u-\!\tilde{\epsilon}_{mn}^{s}
(0)\right)+\delta\left(\epsilon_F-\!(1-\!\beta)\!u-\!\tilde{\epsilon}_{mn}^{s}
(0)\right)+ u\frac{d\beta}{du}\right. \nonumber\\
\!\!\!\!
\left.\times\left(\delta\left(\epsilon_F + \beta u - \tilde{\epsilon}_{mn}^{s}
(0)\right)-\delta\left(\epsilon_F-(1-\beta)u-\tilde{\epsilon}_{mn}^{s}
(0)\right)
\right)\right\},\nonumber
\end{eqnarray}
\begin{eqnarray}
\!\!\!\!\!\!\!\!\!\Upsilon(eU) \approx \frac{F_0}{2H_0} 
 \sum_{s,n,m}\tilde{\epsilon}_{mn}^{s}(0) \left(\frac{\partial\tilde{\epsilon}_{mn}^{s}(\eta)}
{\partial\eta}\right)_{\eta=0}\\
\!\!\!\times\left\{
\frac{\theta(\epsilon_F+\!\beta u\!-\tilde{\epsilon}_{mn}^{s}(0))}
{\left(\epsilon_F\!+\beta\!u-\tilde{\epsilon}_{mn}^{s}(0)\right)^{1/2}} 
+\frac{\theta(\epsilon_F\!-(1\!-\!\beta)u-\!\tilde{\epsilon}_{mn}^{s}(0))}
{\left(\epsilon_F\!-(1\!-\!\beta)u-\!\tilde{\epsilon}_{mn}^{s}(0)\right)^{1/2}} \right\}, \nonumber
\end{eqnarray}
   Note that
 the magnetconductance and magnetotension coefficients contain
 the first 
power of $\partial \tilde{\epsilon}_{mn}^s(\eta )/\partial\eta$.
Therefore   they are  {\em exactly} zero,
due to cancellation of 
terms with opposite $m, s$:
\begin{equation}
\sigma = 0;\:\: \Upsilon = 0.
\end{equation}
This means, that the magnetoconductance and magnetotension in a metal
quantum contact are the effects of   second order in $\eta \ll 1$.
They would  thus  appear as numerically small, extremely narrow peaks in voltage dependence of
the corresponding functions, and cannot be in a satisfactory way 
investigated in our simple model, neglecting the effects of finite 
temperature and scattering. 

\begin{figure}
\epsfxsize=8 cm
\epsfbox{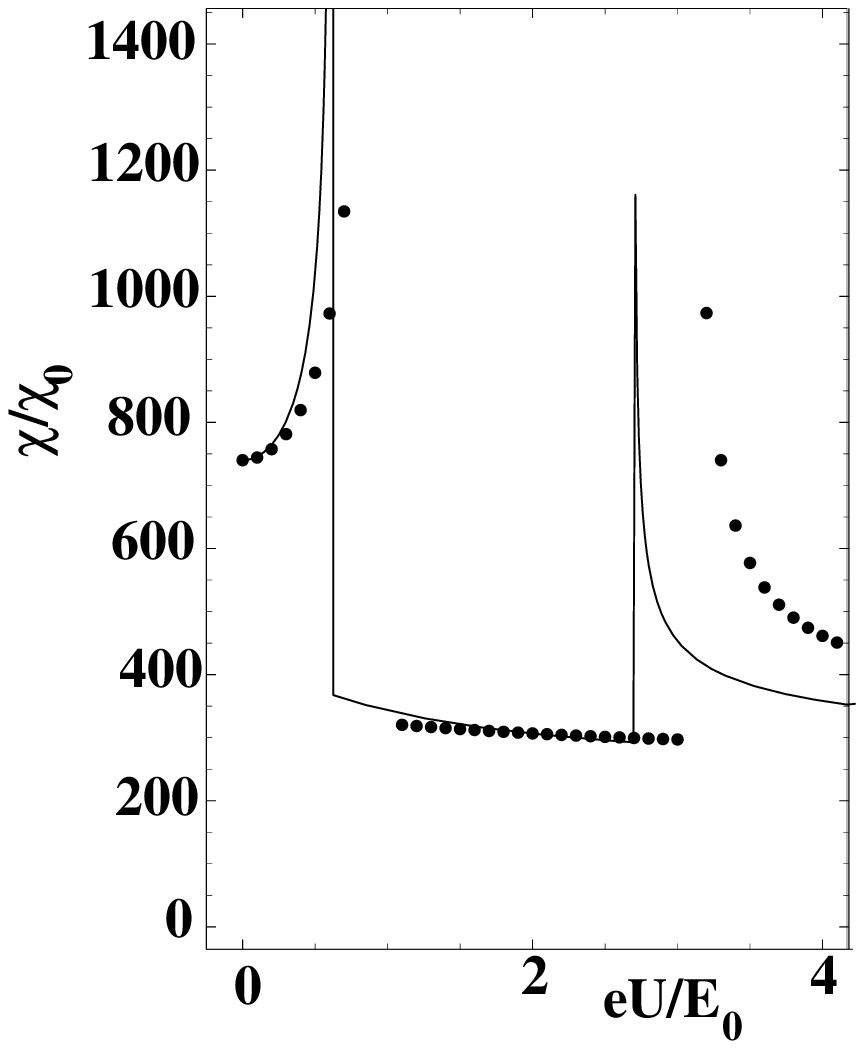}
 \caption{\mbox{Magnetic susceptibility vs. applied voltage in a} 
\mbox{nanowire of round cross section.
 The  unit $\chi_0 = m_e\mu_B/(m^*H_0)$,}
where $H_0 = 8\hbar c/(d^2e)$. We chose $g=2,\:m^*=m_e.$}\label{fig4}
\end{figure}

On the contrary, the magnetic susceptibility  contains the
 second power of $\partial \tilde{\epsilon}_{mn}^s(\eta )/\partial\eta$, and is thus nonzero:
 \begin{eqnarray}
\!\!\!\!\!\!\!\!\!\!\!\!\!\chi(eU) \approx 
\frac{m_e\mu_B}{2m^*H_0}
\sum_{s,m,n}\left( 
\frac{\partial\tilde{\epsilon}_{mn}^s(\eta)}{\partial\eta} \right)^2_{\eta=0} \\
\!\!\!\!\!\!\!\times\left\{\frac{\theta\left(\epsilon_F+\beta u-\tilde{\epsilon}
_{mn}^s(0)\right)}
{(\epsilon_F+\beta u-\tilde{\epsilon}_{mn}^s(0))^{1/2}}  
 +  \frac{\theta\left(\epsilon_F-(1-\beta)u-\tilde{\epsilon}_{mn}^s(0)
\right)}{(\epsilon_F-(1-\beta)u-\tilde{\epsilon}_{mn}^s(0))^{1/2}}
\right\}.\nonumber
\end{eqnarray}
It demonstrates a series of inverse
square root singularities at the same values of driving voltage, as
the features of differential 
conductance and force fluctuations (see Fig.~\ref{fig4}). The features
of $\chi(eU)$ are better pronounced than those of the former coefficients,
which could outweight 
the small magnitude of the effect and make measurements of magnetic susceptibility of a metal point contact  
a more sensitive tool for investigation of electronic density and
potential redistribution in metal point contacts.

  In conclusion, using a simple model,
we showed that finite driving voltage can lead to 
mechanical force fluctuations and singularities of
magnetic susceptibility  in metal quantum contacts. The mechanism of these
effects is voltage-induced nonequilibrium redistribution of electrons over
quasi-1D subbands in the contact. On the other hand, magnetoconductance and
magnetotension coefficients are shown to be exactly zero, 
and the corresponding effects to be at least of order $(H/H_0)^2$, where
$H_0 \approx 10^3$ T in a nanometer size contact.

Experimental investigation of the predicted
effects would clarify the role played by electronic subsystem in 
behavior of metal quantum contacts.

 I am grateful to  I. Affleck, E. Bogachek, A. Bratkovsky, and S. Rashkeev 
for helpful discussions.

\end{document}